\documentclass[useAMS,usenatbib,usegraphicx,A4]{mn2e}
\usepackage{graphicx}
\usepackage{hyperref}
\usepackage{amsmath,amssymb}
\include{Tex-definitions}

\title[Radio structure of 3C~316]{The radio structure of 3C 316, a galaxy with double-peaked narrow optical emission lines}

\author[An et al.]{
T. An$^{1,3,8}$\thanks{E-mail: antao@shao.ac.cn},  
Z. Paragi$^{2}$, S. Frey$^{4}$, T. Xiao$^{1,9}$\thanks{LAMOST Fellow}, W.A. Baan$^{1,3}$, S. Komossa$^{5}$,  
\newauthor  
K.\'E. Gab\'anyi$^{6}$,  Y.-H., Xu$^{7}$, X.-Y. Hong$^{1,8}$
\\ 
$^{1}$Shanghai Astronomical Observatory, Chinese Academy of Sciences, 80 Nandan Road, 200030 Shanghai, P.R. China\\
$^{2}$Joint Institute for VLBI in Europe, Postbus 2, 7990 AA Dwingeloo, The Netherlands\\
$^{3}$Netherlands Institute for Radio Astronomy (ASTRON), Postbus 2, 7990 AA Dwingeloo, The Netherlands\\
$^{4}$F\"OMI Satellite Geodetic Observatory, P.O. Box 585, H-1592 Budapest, Hungary\\
$^{5}$Max-Planck-Institut f\"{u}r Radioastronomie, Auf dem H\"{u}gel 69, 53121 Bonn, Germany \\
$^{6}$Konkoly Observatory, MTA Research Centre for Astronomy and Earth Sciences, P.O. Box 67, H-1525 Budapest, Hungary \\
$^{7}$Yunnan Astronomical Observatory, Chinese Academy of Sciences, Kunming, China \\
$^{8}$Key Laboratory of Radio Astronomy, Chinese Academy of Sciences, P.R. China\\
$^{9}$Key Laboratory for Research in Galaxies and Cosmology, Shanghai, 200030, P.R. China
}

\begin{document}

\date{Accepted 2013 May 3. Received 2013 May 1; in original form 2013 March 27}
\pagerange{\pageref{firstpage}--\pageref{lastpage}} \pubyear{2013}

\maketitle
\label{firstpage}

\begin{abstract}
The galaxy 3C\,316 is the brightest in the radio band among the optically-selected candidates exhibiting double-peaked narrow optical emission lines. Observations with the Very Large Array (VLA), Multi-Element
Remotely Linked Interferometer Network (e-MERLIN), and the European VLBI Network (EVN) at 5\,GHz have been used to study the radio structure of the source in order to determine the nature of the nuclear components and to determine the presence of radio cores.  The e-MERLIN image of 3C~316 reveals a collimated coherent east-west emission structure with a total extent of about 3~kpc. The EVN image shows seven discrete compact knots on an S-shaped line. However, none of these knots could be unambiguously identified as an AGN core. The observations suggest that the majority of the radio structure belongs to a powerful radio AGN, whose physical size and radio spectrum classify it as a compact steep-spectrum source. Given the complex radio structure with radio blobs and knots, the possibility of a kpc-separation dual AGN cannot be excluded if the secondary is either a naked core or radio quiet.

\end{abstract}
\begin{keywords}
galaxies: active -- galaxies: jets -- galaxies: individual: 3C~316 -- radio continuum: galaxies -- galaxies: ISM
\end{keywords}

\section{Introduction}
\label{sec1}

In the framework of the hierarchical formation models of galaxies, mergers play an important role in building massive galaxies, transporting gas toward the galactic centres, initiating bursts of nuclear star formation, and/or triggering the activity of the central supermassive black holes (SMBHs) \citep[e.g.][]{Spr05}. The spatial structure of merging systems and emission characteristics of the structural component provide important insights into physical processes. In particular, the presence of any distinct compact nuclear components may indicate that the merging galaxies contained SMBHs in their nuclei.  In addition, the formation of a supermassive binary black hole \citep[SMBBHs;][]{Beg80} in close Keplerian orbits and in spiral times less than the Hubble time \citep{Yu02} provides a physical laboratory for investigating black hole coalescence, the production of gravitational waves, and other interesting astrophysics \citep[reviewed by e.g.][]{MM05,Kom06}. 

Observations of closely spaced nuclear components on pc and sub-pc scales are severely constrained by the limited angular resolution provided by optical and X-ray techniques and require imaging in the radio with high-sensitivity Very Long Baseline Interferometry (VLBI). However, for this to work, an essential and very strict requirement is that both nuclei are active radio emitters. Until now, only one chance detection has been made of a closely-separated SMBBHs in the radio galaxy 0402+379 \citep{Rodr06}, while only a few larger separation pairs have been identified in X-rays \citep[e.g. ][]{Kom03,Fab11}. A blind search for dual flat-spectrum cores in the data archives of the Very Long Baseline Array (NRAO - VLBA) also failed to identify any other pc-scale-separation nuclear pairs among 3114 sources \citep{Burk11}. 
Considering the difficulty in finding closely-bound (pc-scale) binary BHs, observation of AGN sources at earlier stages of mergers and prior to the nuclear coalescence may be more profitable with typical AGN separations of a few kpc. 

In the optical, double-peaked narrow emission lines (DPNL) is thought as a possible signature of dual AGN. Two velocity components originating from two distinct narrow-line regions (NLRs) have different line-of-sight velocities due to orbital motion. Recent systematic searches for dual AGN showing double-peaked [O\,{\sc III}] line profiles have been made using the Sloan Digital Sky Survey (SDSS) database \citep[e.g.][]{Zhou04,XK09,Wang09,Smi10,Liu10a,Shen11} resulting in an initial sample of about 200 DPNL AGN. However, follow-up high-spatial-resolution imaging and long-slit, integral-field spectroscopic observations found that the majority of these candidates are not genuine dual objects \citep{Liu10b,Shen11,Fu12} and their DPNL line profiles may be explained with peculiar NLR kinematics. An even larger sample is needed for a better understanding of the galaxy merger processes and the nature of DPNL systems \citep{Dot12}. VLBI, the currently highest resolution imaging technique, is a powerful tool to clarify the AGN duality.
However only a few DPNL dual AGN candidates have been observed with VLBI so far. 
For instance,
recent observations with the European VLBI Network (EVN) of one of these DPNL sources, the AGN J1425+3231 showing a double-peaked [O\,{\sc III}] line profile in its SDSS spectrum \citep{Peng11}, successfully detected two compact radio components with a projected linear separation of $\sim$2.6~kpc \citep{Frey12} .  The estimated high brightness temperatures ($>$$10^7$K) and the small sizes (7.4~pc and 2.7~pc) suggest dual AGN. 

In this first of a systematic study of dual AGN candidates, radio interferometry imaging techniques are employed for the source 3C\,316 (SDSS J151656.59+183021.5) in order to investigate its DPNL character. 3C\,316 resides in a 19 mag host galaxy at redshift  $z$=0.5795 \citep{GW08}. It has a flux density of 1.3~Jy at 1.4~GHz \citep{Edge59,WB92}, and is the most radio-loud AGN in the DPNL AGN sample \citep{Smi10}. 
Although this source is quite bright in both the optical and radio regimes, it has no radio interferometric image published in the literature and has not been observed with VLBI. The source 3C~316 may indeed contain dual AGN, or  alternatively its DPNL nature is attributed to peculiar NLR kinematics resulting from bipolar jets/outflows \citep[e.g.][]{Hec81,Hec84} or the rotation of disk-like NLRs \citep[e.g.][]{GH05}, or from a single AGN ionizing the ISM of two galaxies \citep{XK09}.

In this paper, we present the radio images of 3C\,316 obtained with the EVN, e-MERLIN and VLA at 5~GHz. The SDSS optical spectrum is also analyzed in order to supplement the radio data.
Section~\ref{sec2} describes the radio observations and data reduction. The radio images and the SDSS spectrum of 3C\,316 are presented in Section~\ref{sec3}. In Section~\ref{sec4}, the radio properties of 3C\,316 are discussed.  The results are summarized in Section~\ref{sec5}. In the cosmological model with $H_0=73$\,km\,s$^{-1}$\,Mpc$^{-1}$, $\Omega_\mathrm{M}=0.27$ and $\Omega_\mathrm{\Lambda}=0.73$ \citep{Spe07} used throughout this paper, 1\,mas angular size corresponds to 6.38\,pc projected linear size at $z$=0.5795 \citep{Wri06}.

\section{Observational Data}
\label{sec2}

\subsection{The EVN data}
\label{sec2-1}

3C\,316 was observed with the EVN at 5~GHz in e-VLBI mode on 2011 March 22 (project code: EA047). Nine telescopes (Effelsberg, Jodrell Bank Mk~II, Medicina, Onsala, Toru\'n, Westerbork, Yebes, Shanghai, Hartebeesthoek) participated in the observations. The maximum data rate per station was 1024\,Mbps for a total bandwidth of 256~MHz. The observations were carried out in phase-referencing mode. The bright and compact quasar J1516+1932 ($\sim$0.8\,Jy; $1 \fdg 03$ away from 3C\,316) was used as the phase-reference calibrator. The observations lasted 2\,h and used a 'calibrator (1.5\,min) -- target (3.5\,min) -- calibrator (1.5\,min)' sequence, resulting in an on-target time of 1.2\,h. The data were correlated in real time with the EVN Mk~IV Data Processor at the Joint Institute for VLBI in Europe (JIVE, Dwingeloo, the Netherlands) with an integration time of 2\,seconds.
 
The data reduction followed the online EVN guide\footnote{\url{http://www.evlbi.org/user\_guide/guide/userguide.html}}. Calibrations were made using the NRAO Astronomical Image Processing System ({\sc AIPS}) software package \citep{Gre90}. The visibility amplitudes were calibrated using the system temperatures measured during the observations and the gain curves of each telescope. The calibrated data were averaged in each sub-band and split into single-source files. A few iterations of phase-only self-calibration of the 3C~316 data were made in the Caltech software package {\sc Difmap} \citep{She97} to remove residual phase errors. The Caltech software package program {\sc MAPPLOT} was used to produce the final images. 

\subsection{The e-MERLIN data}
\label{sec2-2}

The 5-GHz observations of 3C~316 with the Multi-Element Remotely Linked Interferometer Network (e-MERLIN) were carried out at two epochs separated by nearly 3 days (2011 April 14 UT~20:00--April 15 UT~09:00, and 2011 April 17 UT~19:00--April 18 UT~09:00). Five telescopes (Jodrell Bank Mk~II, Knockin, Defford, Pickmere, Darnhall) participated in the observations which provided a maximum baseline length of 100~km. The observations were made in dual-polarization mode with a total bandwidth of 512~MHz. The on-source time for 3C~316 was 9.1~hours at each date. Three calibrator sources were observed: J1516+1932, 3C~286 and OQ~208. Because 3C~286 is resolved with the e-MERLIN, OQ~208 was used to calibrate the flux density scale. The long-term 5-GHz light curves of OQ~208 monitored by the University of Michigan Radio Observatory  \citep[e.g.][]{aller85} are rather stable from the middle of 1990s to 2010 \citep{Wu13} and gave an estimated flux density of 2.46~Jy. The complex gain solutions were derived from the calibrators OQ~208 and J1516+1932, and then applied to the whole data in {\sc AIPS}. The flux density calibration was verified by comparing the flux density of J1516+1932 independently observed with the e-MERLIN and the EVN. After calibration, the flux density of J1516+1932 derived from the e-MERLIN data is 0.742~Jy at the central frequency of 4.67~GHz, which is in excellent agreement with the EVN value (0.749~Jy at 4.99~GHz) just three weeks earlier. The e-MERLIN data sets at the two individual epochs were calibrated separately and then combined. Further phase and amplitude self-calibration of the 3C~316 data were done in {\sc Difmap}. The uncertainty of the measured flux densities is estimated to be $\lesssim$5 per cent.

\subsection{The archival VLA data}
\label{sec2-3}

The VLA data for 3C~316 were acquired from the NRAO archive (project code: AH167). These observations were conducted at 4.85~GHz in the A-array configuration on 1984 December 8. Snapshot mode was used in the experiment and 3C~316 was observed during a single scan of $\sim$1.2 min. The data were recorded in two adjacent 50-MHz intermediate frequency channels (IFs). For the sake of comparison with the EVN and e-MERLIN observations, the ($u,v,w$) coordinates of the visibilities were first recalculated by converting source coordinates from B1950 to J2000 using the {\sc AIPS} task {\sc UVFIX}. Then the data calibration was carried out in {\sc AIPS} following the standard procedure described in the AIPS Cookbook. The flux density scale was set by the observation of 3C~286 assuming a total flux density of 7.49~Jy at 4.85~GHz. A number of bright and compact AGN targets were observed during the project in order to measure the antenna-based gains and phase solutions to be applied to the whole dataset. The calibrated visibility data of 3C~316 were imported into {\sc Difmap} for self-calibration and imaging.

\subsection{Composite radio data}
\label{sec2-4}

The calibrated VLA and e-MERLIN visibility data sets were concatenated to allow for creating an intermediate-resolution image. Apart from the conversion from the B1950 to the J2000 coordinate system, the phase centre of the VLA data was shifted to coincide with that of the e-MERLIN data using the {\sc AIPS} task {\sc UVFIX}. The visibility amplitudes of the two data sets at common VLA and e-MERLIN baseline ranges were found to be consistent despite the different observing epochs. A few iterations of phase-only self-calibration eliminated the minor misalignment errors and residual phase errors.  

The EVN and e-MERLIN data were combined in a similar way. Both data sets had the same pointing centre and used J2000 coordinates. Only the possible amplitude scaling difference between the two sets of visibility data required caution. The total flux density of 3C~316 is dominated by extended emission (see Section~\ref{sec3}). When inspecting the e-MERLIN visibility amplitudes as a function of the projected baseline length, we found that the variation pattern of the amplitude is roughly in conjunction with that of the EVN data on the shortest baselines. An iteration of amplitude and phase self-calibration was made to eliminate the residual error in the amplitude scale of the composite data set. 

\subsection{SDSS data}
\label{sec2-5}

The optical spectroscopic data of 3C~316 were obtained from the SDSS Data Release 7\footnote{\url{http://www.sdss.org/dr7/}} \citep[DR7,][]{Aba09}. The data were already calibrated in the SDSS spectroscopic pipeline. The final spectrum covers the observed-frame wavelength range 3800--9200~\AA{} (Fig.~\ref{fig2}).

\section{Results}
\label{sec3}

\subsection{Radio morphologies}

\begin{figure*}
\includegraphics[width=5.6cm,height=5.5cm]{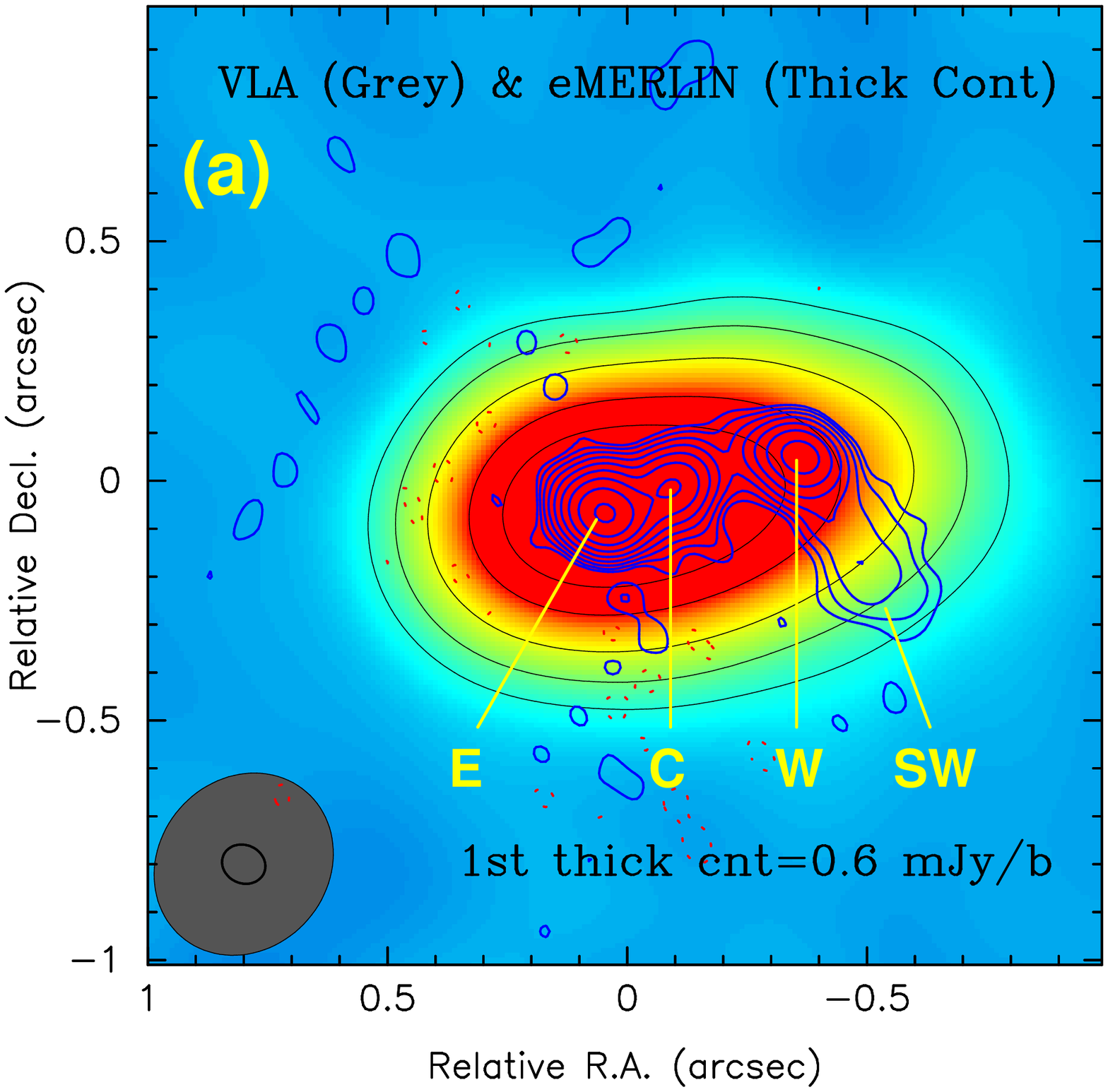} \hspace{2mm}
\includegraphics[width=5.6cm,height=5.5cm]{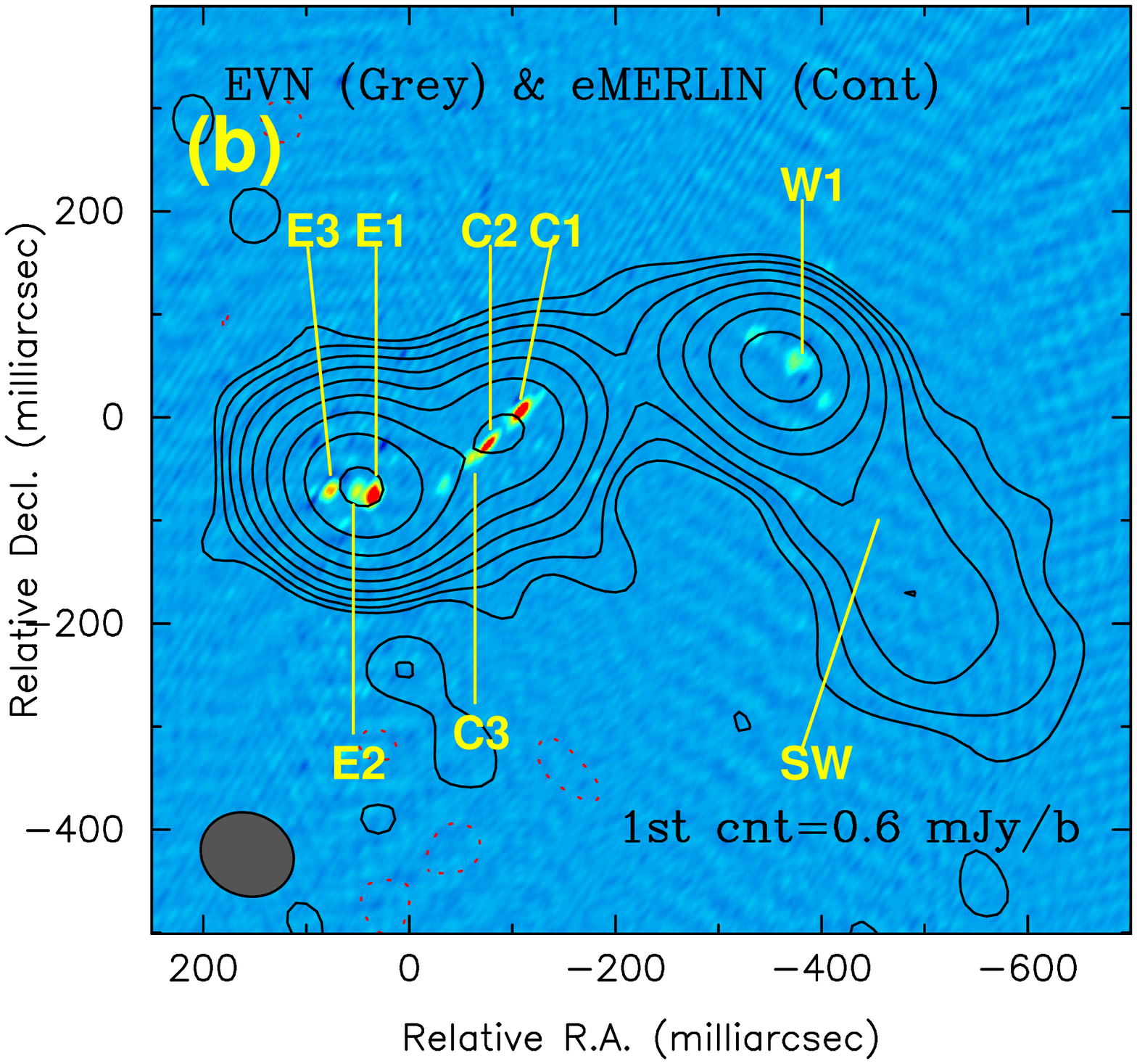} \hspace{2mm}
\includegraphics[width=5.6cm,height=5.5cm]{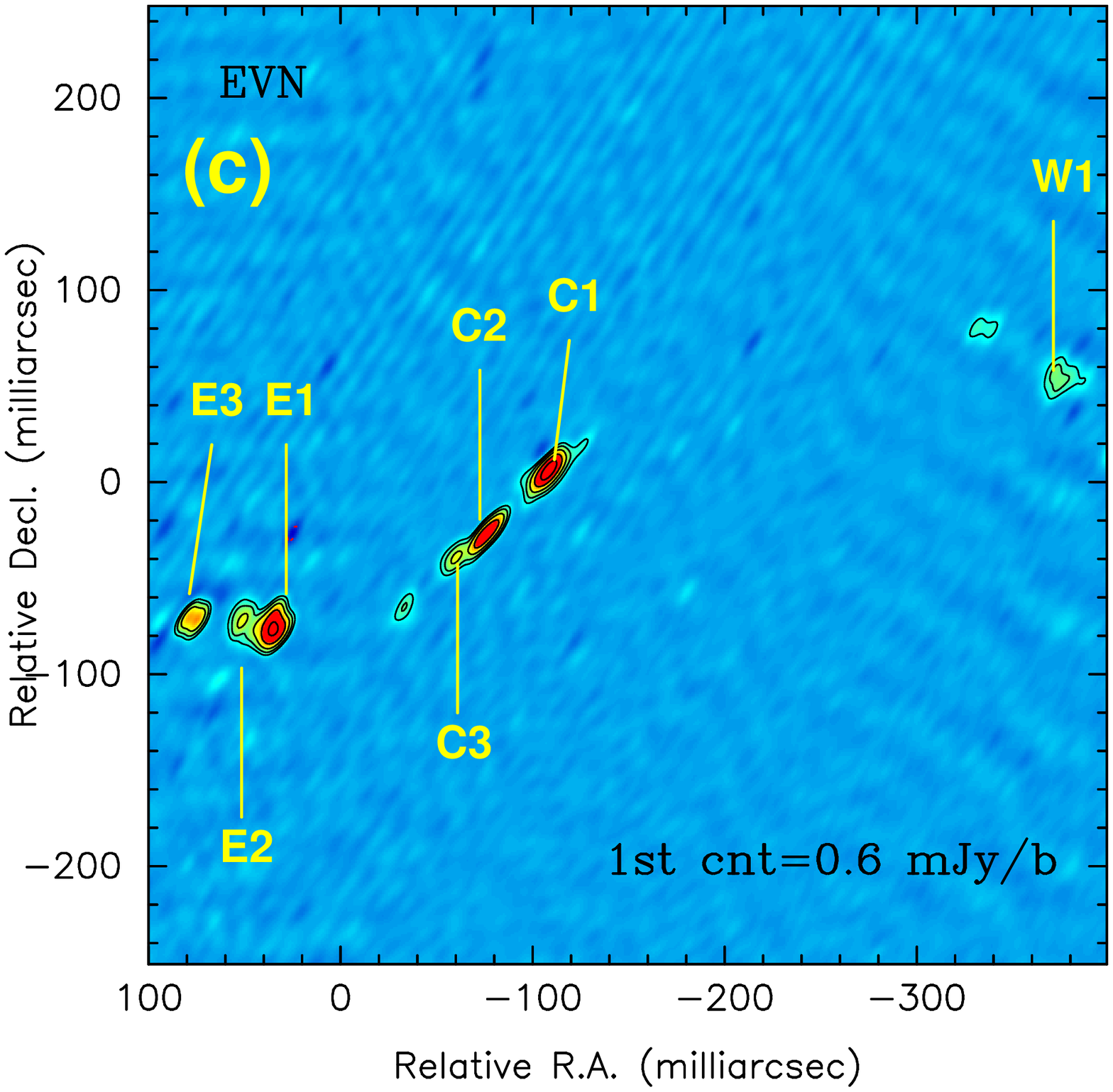}
\caption{(a): The 5-GHz VLA image of 3C~316 (colour scale and thin contours) overlaid with the e-MERLIN image (thick contours). The image centre is at R.A.(J2000)=15:16:56.597, Dec.(J2000)=18:30:21.560. The restoring beam of the VLA image is 400$\times$354 mas$^2$, P.A.=$-$40.8\degr. The rms noise is 0.64~mJy beam$^{-1}$. The thin contours represent the intensities of 2.0~mJy beam$^{-1}$ $\times$(1,2,4,8,12). The restoring beam of the e-MERLIN image is 92.3$\times$79.7 mas$^2$, P.A.=66.3\degr. The rms noise is 0.15~mJy beam$^{-1}$. The thick contours are at 0.6~mJy beam$^{-1}$ $\times$(1,2,4,8,16,32,64,128,256).
(b): The 5-GHz EVN image (colour scale) of 3C~316 overlaid with the e-MERLIN image (contours). The image centre is at R.A.(J2000)=15:16:56.597, Dec.(J2000)=18:30:21.560. The restoring beam of the EVN image is 14.1$\times$4.9 mas$^2$, P.A.=$-$31.8\degr. The rms noise is 0.05~mJy beam$^{-1}$. The peak intensity is 14.5~mJy beam$^{-1}$ at the component E1 (R.A.(J2000)=15:16:56.590, Dec.(J2000)=18:30:21.567). The parameters of the e-MERLIN image are the same as for Fig.~\ref{fig1}a. The peak of 173.0~mJy beam$^{-1}$ is located between E1 and E2, at R.A.(J2000)=15:16:56.601, Dec.(J2000)=18:30:21.490. 
(c): A tapered EVN image of 3C~316. A circular Gaussian restoring beam of 35$\times$35 mas$^2$ is used. The rms noise is 0.4~mJy beam$^{-1}$. The contours are at 1.6~mJy beam$^{-1}$ $\times$(-1,1,1.414,2,4,8,16,32).}
\label{fig1}
\end{figure*}

The VLA image of 3C~316 (colour scale) is shown in Figure \ref{fig1}a. The source is resolved and elongated in the east--west direction. The total flux density is dominated by the extended emission structure. The outermost boundary is not accurately constrained because it is much sensitive to the dynamic range of the image. The integrated flux density measured over the source is 393.3~mJy, consistent with the single-dish flux density (414$\pm$56~mJy) measured at Green Bank \citep{GC91}. Given that prominent variability is not reported for 3C\,316, no larger-scale radio emission is resolved out because of missing short spacings. 

The e-MERLIN image (contours in Fig.~\ref{fig1}a), with a resolution of $0\farcs09$$\times$$0\farcs08$, resolves the source into a connected triple structure with an eastern (E), a central (C), and a western (W) component. The e-MERLIN structure is centered on the central part of the VLA structure. Because the shortest baseline of e-MERLIN is about 50~k$\lambda$, as compared to 10~k$\lambda$ for the VLA, the major VLA structure is sampled by e-MERLIN. The estimate of the integrated flux density is $388.4$$\pm$$19.4$~mJy in the e-MERLIN image. The difference between the VLA and e-MERLIN integrated flux density is only 4.9~mJy, within the uncertainty of the flux density measurements. Possibly there is a more extended halo, which is resolved out in the e-MERLIN image. An interesting feature revealed by the e-MERLIN image is a collimated and extended feature (labelled as SW) to the southwest of the western component W, which cannot be discerned in the VLA image. This SW feature extends to a distance of about $0.5\arcsec$ (or $\sim$3~kpc) and has a peak intensity of about 4.8 mJy beam$^{-1}$, corresponding to $\sim$30 times the rms noise.

An intermediate-resolution image using the VLA and e-MERLIN data is more sensitive to extended structure but looks much like the e-MERLIN-only image (not shown here).  Except for the triple structure already seen in the e-MERLIN image, the composite image also shows some diffuse emission to the north and southwest of the main radio structure. The northern diffuse structure disappears in the e-MERLIN-only image. 

The full-resolution EVN image is displayed in colour scale in Figure \ref{fig1}b. The restoring beam is 14.1$\times$4.9 mas$^2$, P.A.=$-$31.8\degr. 
A tapered EVN image is displayed in Figure \ref{fig1}c using {\sc IMAGR} in {\sc AIPS} with robust weighting (robust=1). The visibility data have been tapered with a Gaussian value of 0.5 at the projected $uv$-distance of 15~M$\lambda$, to decrease the dominance of the longest baselines in the data. The main features in the e-MERLIN image are resolved into seven discrete knots in the EVN image, which highlights the ridge line of the contours of the e-MERLIN image. These bright knots cluster into three groups associated with the three components in the e-MERLIN image: the eastern (E1, E2 and E3), central (C1, C2, C3) and western groups (W1 and two weaker knots). The brightest VLBI component E1 coincides with the e-MERLIN peak. The connecting lines of the emission peaks of these VLBI components resemble an S-shaped trajectory. A sharp curvature occurs around E1 and between C1 and W1, as manifested in the contours in the e-MERLIN image. 
\begin{table}
\caption{Model fitting results of compact VLBI components.} 
\label{tab1}
\vspace{-3mm}
\begin{center}
\begin{tabular}{cccccc}
\hline
Comp. &$R_{\rm sep}$ & P.A.  & $\theta_{\rm maj}$, $\theta_{\rm min}$ & $S_{\rm int}$  & $T_{\rm b}$     \\
      & mas    & $\degr$  & mas, mas     & mJy   & $10^6$~K  \\
(1)   & (2)    & (3)      & (4)          & (5)   & (6)  \\
\hline
E3    & 104.8  & 132.7    & 11.4, 8.5    &  20.0 & 15.9 \\   
E2    &  88.7  & 145.2    & 21.9, 11.3   &  25.8 &  8.0 \\   
E1    &  84.0  & 155.2    & 11.7, 9.4    &  40.4 & 28.3 \\
C3    &  69.8  & $-$124.9 & 11.8, 4.6    &   7.4 & 10.5 \\   
C2    &  81.2  & $-$108.9 & 14.0, 2.4    &  16.5 & 37.8 \\
C1    & 108.7  & $-$86.7  & 10.8, 4.4    &  26.2 & 42.4 \\
W1    & 376.6  & $-$82.2  & 18.8, 11.0   &  13.6 &  5.1 \\   
\hline
\end{tabular} \\
\end{center}
(1): the labels of the VLBI components used in Fig.~\ref{fig1}c; 
(2) and (3): the relative separation and position angle with respect to the image centre at RA=15:16:56.597 and Dec=18:30:21.560, respectively;  
(4): the deconvolved size (major and minor axes) of the elliptical Gaussians (FWHM); 
(5): the integrated flux density of the components; 
(6): the brightness temperature calculated from the observables.
\end{table}

To quantitatively investigate the properties of the compact VLBI components, we fitted the seven bright components with Gaussian brightness distribution models using the {\sc Difmap} task {\sc modelfit}. The parameters of the fitted elliptical Gaussian model components are listed in Table~\ref{tab1}. The sum of the flux densities of all components is 150~mJy, which is about 50 per cent higher than the total CLEANed flux density in the EVN image, because of model fitting done in the visibility domain. Column~6 in Table~\ref{tab1} gives the brightness temperatures ($T_{\rm b}$) of the components in the source rest frame, calculated in the units of K using the equation \citep{KO88}: 
\begin{equation}
T_{\rm b} = 1.22\times 10^{12} \frac{S_\nu}{\theta_{\rm maj}\theta_{\rm min} \nu^2} (1+z), 
\end{equation} 
where the component flux density $S_\nu$ is in the units of Jy, the source size $\theta_{\rm maj}$ and $\theta_{\rm min}$ (the Gaussian major and minor axes of the full width at half maximum, FWHM) in units of mas, and the observing frequency $\nu$ in units of GHz. In the case of 3C\,316, the brightness temperatures of the VLBI components range from $5.1$$\times$$10^6$~K to $42.4$$\times$$10^6$~K.

\subsection{Optical signature}

\begin{figure*}
\begin{center}
\includegraphics[width=0.8\textwidth]{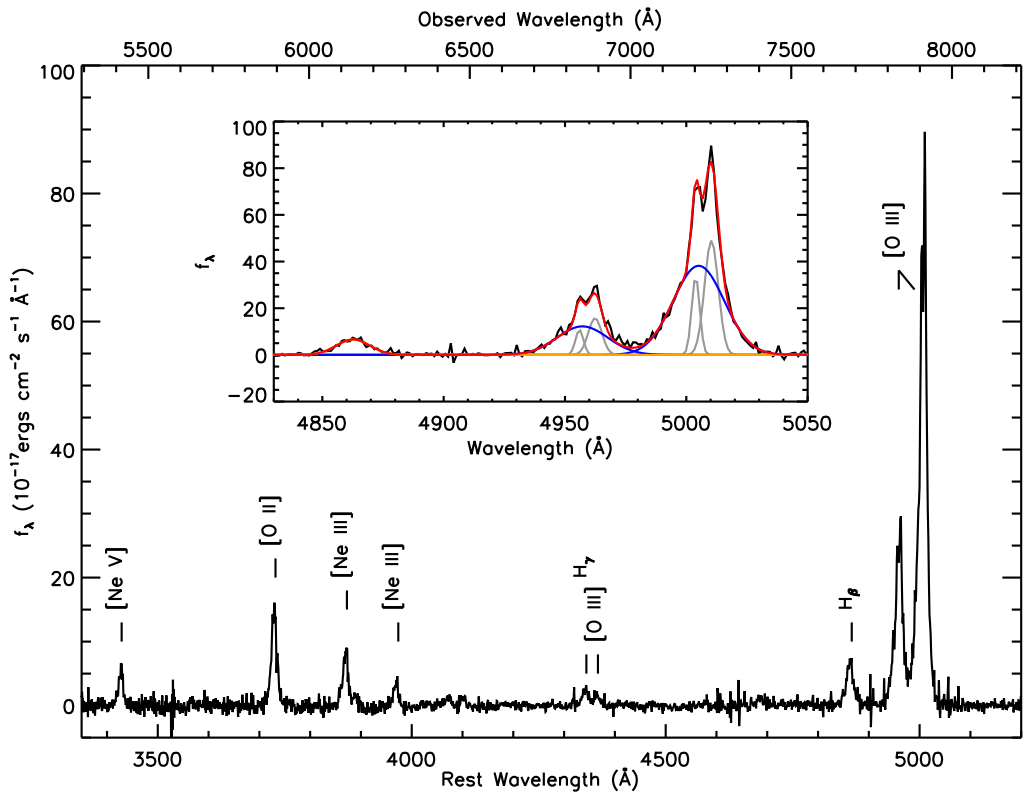}
\end{center}
\vspace{-3mm}
\caption{SDSS spectrum of 3C~316. The spectrum has been corrected Galactic extinction, and shifted to the source rest frame according to the line centre of H${\beta}$. The [O\,{\sc III}]$\lambda$4959, $\lambda$5007 and H$\beta$ spectra are highlighted in the inset. 
The blue curves represent the Gaussian fits with a broad-velocity and the gray curves denote two narrow-velocity components. The red curve denotes the composite of the Gaussian components.
}
\label{fig2}
\end{figure*}

The archival SDSS spectroscopic data were re-analyzed to study the physical properties of the ionised gas and to identify the  velocity structure of the emission lines using the following reduction procedure.  

The spectrum of 3C~316 was {\it corrected for the Galactic extinction} using the dust extinction map of \citet{SFD98} with an extinction curve of \citet{Fit99} with $R_{\rm V}=3.1$. 
The spectrum was {\it shifted to the source rest frame} using the line centre of the H${\beta}$ line.   

An important analysis step is {\it modeling of the continuum} using starlight and nuclear continuum \citep{Dong05,Xiao12}. The starlight continuum was modeled by a linear combination of a set of synthetic templates, constructed from the simple stellar population library \citep{BC03}, by using a method based on the Ensemble Learning Independent Component Analysis. The details of template construction and starlight continuum model are given in \citet{Lu06} and \citet{Zhou06}. In brief, the templates were broadened by convolving with a Gaussian and were shifted to match the stellar velocity dispersion of the galaxy. Then the template spectra are reddened assuming the extinction curve of the Small Magellanic Cloud \citep{Pre84} with a varying E(B-V) between 0 and 1. The nuclear continuum was modeled with a power-law ($\lambda^{-\alpha}$), in which the spectral index was assumed to be $\alpha$$=$$-1.7$ \citep{Fra96}. Bad pixels in the spectrum were flagged out following the SDSS pipeline, and the emission lines were excluded in the power-law fit. The continuum fits were also tested with the $F$-test by including and excluding the nuclear component, which showed that the inclusion of the power-law nuclear continuum greatly improved the continuum fit to a significance level of 99 per cent. A composite of the stellar and nuclear continua was then subtracted from the original spectrum. Fig.~\ref{fig2}b shows the continuum subtracted spectrum.
 
The continuum-subtracted spectrum was used for {\it fitting the line components}. The [O\,{\sc III}] $\lambda$5007 line is the most prominent. The line shows a distinctive double-peaked profile and a broad blue-shifted shoulder, to be fitted with three Gaussian components. The fitted [O\,{\sc III}] line components are presented in Table \ref{tab2}. The line centre of the broad shoulder lies between the peaks of the two narrow components, but relatively closer to the blue narrow component. The [O\,{\sc III}] $\lambda$4959 line is the second strongest line in Fig.~\ref{fig2} and has been fitted in a same manner as the $\lambda$5007 line.
Differently from the [O\,{\sc III}] $\lambda$5007 and $\lambda$4959 lines, the other narrow lines ([O\,{\sc II}], H$\beta$, [Ne\,{\sc III} ], and [Ne\,{\sc V}]) do not show double-peaked profiles, possibly because of blending of multiple velocity components. A single Gaussian was used to fit these weaker lines. All parameters derived from Gaussian fits are given in Table~\ref{tab2}.

\begin{table*}
\caption{Optical line properties of 3C~316.}
\begin{tabular}{ccccccccc}
\hline
Line      &   H$\beta$ & [O {\sc III}] $\lambda$4959 & [O {\sc III}] $\lambda$5007 & [O {\sc II}] & [Ne {\sc III}] $\lambda$3870 & [Ne {\sc III}] $\lambda$3968  & [Ne {\sc V}]\\
\hline
\multicolumn{7}{c}{Blue Broad Component}   \\
$\lambda_c$ (\AA) & 4862.68$\pm$0.25 & 4957.18 & 5005.10$\pm$0.12 & 3731.77$\pm$0.19 & 3869.52$\pm$0.21 & 3968.88$\pm$0.41 & 3427.16$\pm$0.29 \\
V$_c$ (km s$^{-1}$) & 0 & $-$187.8 & $-$188.0 & $+$36.6 & $-$49.5 & $-$2.5 & $+$58.1 \\
$\Delta V$ (km s$^{-1}$) & 426.31$\pm$15.10 & 630.72 & 630.72$\pm$6.64  & 458.33$\pm$15.45 & 399.7 & 346.8 & 399.7 \\
L ($10^{42}$~erg~s$^{-1}$) & 2.33$\pm$0.07 & 6.52 & 20.56$\pm$0.19 & 4.03$\pm$0.12 & 2.21$\pm$0.07 & 0.81$\pm$0.01 & 1.28$\pm$0.01 \\
\hline
\multicolumn{7}{c}{Blue Narrow Component}   \\
$\lambda_c$ (\AA) &  & 4955.96 & 5003.85$\pm$0.11 \\
V$_c$ (km s$^{-1}$) & & $-$262.2 & $-$262.3 \\
$\Delta V$ (km s$^{-1}$) &  & 99.08  & 99.08$\pm$5.70  \\
L ($10^{42}$~erg~s$^{-1}$) & & 0.90 & 2.82$\pm$0.21  \\
\hline
\multicolumn{7}{c}{Red Narrow Component}   \\
$\lambda_c$ (\AA) &  & 4962.38 & 5010.34$\pm$0.10 \\
V$_c$ (km s$^{-1}$) & & $+$126.1 & $+$125.9 \\
$\Delta V$ (km s$^{-1}$) &  & 169.54 & 169.54$\pm$6.07 \\
L ($10^{42}$~erg~s$^{-1}$) & &2.25 & 7.10$\pm$0.07\\
\hline
\end{tabular}
\label{tab2}
\end{table*}
\section{Discussion}
\label{sec4}

\subsection{Finding the nucleus}
\label{sec4-1}

The nuclear structure of the radio-loud 3C~316 is of interest because of the double-peaked [O\,{\sc III}] line in its optical spectrum. The seeing in the SDSS optical imaging data is $\sim$1.4\arcsec, which corresponds to $\sim$9~kpc for 3C~316. Since most of the dual AGN candidates remain optically unresolved in SDSS images, observations with higher spatial resolution are necessary to distinguish between the dual AGN and NLR kinematics scenarios.
Maybe only about 10 per cent of double-peaked [O\,{\sc III}] Type-2 AGNs can be interpreted as dual AGN, while the majority is attributed to NLR kinematics \citep{Smi10,Shen11,Fu12}. 
 
The VLBI components of 3C~316 revealed in our EVN image (Fig.~\ref{fig1}b) appear compact at 14~mas$\times$5~mas (or $\sim$90~pc$\times$30~pc) resolution, but they are already partly resolved. The calculated brightness temperatures of $\sim$$10^6$--$10^7$~K (Table~\ref{tab2}) are much higher than expected from the starburst galaxies \citep{Con92}, but significantly lower than the typical $T_{\rm b}$ of Doppler-boosted radio-loud AGN cores \citep{Rea94}.  
Instead, the VLBI features resembles the knots in the radio jets, which can be associated with velocity irregularities in the flow, re-confinement or re-collimation shocks, interactions between jets and the external environment (e.g., clouds), or large-scale instabilities in the flow. Alternatively, the jet knots could be the components that correspond to episodes of enhanced output of the central engine.
Typically, the jet knots in the radio galaxies Pictor~A \citep{Tin08} and in Centaurus~A \citep{TL09} have $T_{\rm b}$$\sim$$10^{5}$--$10^{6}$~K. In Doppler-boosted jets, the $T_{\rm b}$ of jet knots can be as high as $10^{9}$~K \citep[e.g. in 3C~48,][]{An10}, but seldom exceed $10^{10}$~K. Although 3C\,316 does not show any prominent core components, it is possible that there are one or more synchrotron self-absorbed AGNs in the source. The following possible scenarios are available.

{\em Two radio AGN: both with prominent jets.} 
In this scenario, the radio features E1 and W1 may correspond to the AGN cores. The west AGN shows an extended jet to the southwest (feature 'SW'). The kpc-scale separation of two AGN is consistent with those in previously discovered DPNL dual AGN \citep[e.g.][]{Liu10a,Shen11,Fu11}. 
A major difficulty of this interpretation is the bridge (radio feature C) linking the two possible AGN. Radio continuum bridges are not unusual in merging and interacting galaxies, however the compact appearance (at VLA and e-MERLIN resolutions) of the prominent central component C is not analogous to a stretched, diffuse morphology of a 'Taffy' galaxy pair \citep[e.g., the UGC 12914/5 pair,][]{Con93}. Moreover, the high flux density of the component C at 5~GHz (7.9~GHz in the source rest frame) requires a highly-ordered strong magnetic field and a vast amount of cosmic rays to generate the synchrotron radiation. Alternatively, the component C might be the jet component associated with the East AGN. However, there is no reason why the jet of the East AGN should point exactly to the West AGN.  

{\em Two radio AGN: one has a two-sided jet and the other is a 'bare' core.} Different from the first scenario, the major radio structure here is associated with a powerful radio AGN, while the other AGN only shows a naked core as found in 0402+379 \citep{Rodr06}. The possible candidates of AGN cores are the bright  C1 and C2 component separated by about 180 parsec. A simple integrated mass estimate of a binary BH system with such a separation and a (optical) velocity difference of 388.3 km s$^{-1}$ places a lower limit of $M_{int} = 7.8 \times 10^8$ $M_{\odot}$, assuming the orbital plane of the binary BH is perpendicular to the plane of sky. Other orientation would give a larger integrated mass. 
The naked-core AGN, either C1 or C2, has been triggered recently and jets or extended lobes have not been formed yet, or they are intrinsically faint.

{\em A radio-loud/radio-quiet AGN pair.} 
The third scenario is that all radio structure belongs to a radio-loud AGN and that a radio-quiet AGN is hidden somewhere in the system. An example of such a configuration is PKS~0347+05, which consists of a FR~II radio galaxy and a Seyfert 1 AGN, separated by 25~kpc \citep{Tad12}. However, the optical image of 3C~316 has sufficient resolution to confirm this scenario but shows no resolved morphology. In this regard, even if the possibility of two optical AGN confined within about 10~kpc cannot be entirely ruled out, the observed radio structure should belong to a single radio-loud AGN. 

To summarize, the last two scenarios appear possible and the first one appears somewhat unlikely. Based on the brightness temperatures of the 3C~316 knots, a binary AGN cannot be positively identified, but given the complex radio structure, it  cannot exclude either. Multi-frequency observations are needed to identify any true core(s). In the following discussions, the majority of the coherent emission structure is associated with a powerful radio-loud AGN.

\subsection{Radio source structure} 

Three possible models may be used to interpret the radio morphology.

{\em (1) A triple source consisting of a central core (C) and two lobes (E and W).} The non-detection of high $T_{\rm b}$ core invokes a possibility that 3C~316 may contain an obscured and self-absorbed AGN in its centre. The three major components in Fig.~\ref{fig1}b, E, C and W align along an S-shaped configuration approximated as a straight line. E and W lie at nearly symmetric locations with respect to C. A straightforward interpretation of the active nucleus is that it is hidden in the central component C. The projected source size can be estimated from the separation between the two brightest peaks E1 and W1, $0\farcs45$. It gives a projected linear size of $\sim$3~kpc.
 
Because 3C~316 appears to be a Type 2 Seyfert galaxy, the radio jet should have a large viewing angle (close to the plane of the sky) and a physical size almost equal to the projected size.  Therefore, the 3C\,316 morphology (apparent triple structure, small source size, and steep integrated spectrum) resembles a compact double radio structure confined within the AGN narrow-line region \citep[e.g.][]{Ben06}.  The overall projected source size is slightly larger than a typical compact symmetric object \citep[CSOs,][]{PM82,Wil94}, and the source resembles a medium-sized symmetric object \citep[MSO,][]{Rea95,FS96}.  The dimness of the core could be due to synchrotron self-absorption and/or free-free absorption at centimeter wavelengths. The brightest VLBI component C1 with brightness temperature $T_{\rm b}$$\approx$4.2$\times$$10^7$~K and the second brightest C2 with $T_{\rm b}$$\approx$3.8$\times$$10^7$~K, are the most probable candidates of the AGN core. New 1.6-GHz VLBI observation of this galaxy are necessary to provide spectral indices and to distinguish between a flat-spectrum AGN core and steep-spectrum jet knots. 

The intensity ratio of an approaching jet enhanced by Doppler boosting, and the receding counter jet of an intrinsically symmetric source constrains the jet parameters as \citep{LB85}: 
\begin{equation}
R = \left(\frac{1+\beta\cos\theta}{1-\beta\cos\theta}\right)^{\alpha+2},
\label{eq:Iratio}
\end{equation}
where $\beta$ is the jet speed in the units of the speed of light $c$, $\theta$ the inclination angle between the jet axis and the line of sight, $\alpha$ the radio spectral index. Assuming that the brighter component E is the jet component moving toward the observer, and W is the receding one, an intensity ratio $R$=3.1 is derived from the e-MERLIN image. Substituting $R$ into Eq.~\ref{eq:Iratio}, we obtain $\beta\cos\theta$=$0.20$. As discussed above, the optical spectral properties suggest that 3C~316 is a Seyfert 2 galaxy. The dividing orientation angle between Seyfert 1 and 2 sources is $\theta$$\approx$$76\degr$ \citep{Zhang05}. 
This in turn gives an upper limit to the jet speed $\beta$$<$$0.83$ (or $<$0.025~mas~yr$^{-1}$ apparent proper motion), which indicates a moderately relativistic jet. But Eq.~\ref{eq:Iratio} should be used with caution because the jet components may be intrinsically different and their flux densities may be affected by jet--ISM interactions. Furthermore, if the galaxy is not a 'classical' Type 2, but rather obscured by the dusty ISM of its host, no constraints on its orientation can be drawn. Nevertheless, the apparently symmetry of the jet components with respect to the geometric centre C1 suggests that the jet speed could indeed be small.

The lobe advance speed in young radio sources is much smaller than the jet knot speed \citep{An12}, and the dynamic evolution of young radio sources shows a deceleration of the lobe advance during the CSO growth stage \citep{KK06,AB12}. 
Direct determination of such a slow proper motion (0.025~mas~yr$^{-1}$) requires high-resolution VLBI monitoring observations over a long time interval. 
On the other hand, the source size and proper motion limits give a reasonable estimate for the kinematic age of 3C~316, $\sim$$10^4$--$10^5$~yr, which is typical for most compact MSOs or extended CSOs.

The tongue-like SW structure of $\sim$2~kpc in length, which is comparable to the whole source size (Fig.~\ref{fig1}b), looks like a collimated outflow originating from the western lobe W, after the jet loses its momentum at the ISM-IGM boundary or at the boundary of the NLR \citep{AB12}.   Such a feature can be generated when the jet hits the wall in the external medium and flushes out along the pressure gradient or via weak point in the wall \citep{LB86}. 
This structure should be even older since its flow speed cannot be faster than the jet speed and may result from (multiple) episodic activity in the past.

{\em (2) Two-sided jets with large bending angle.}
In this model, the core is located at the component W. The C and E components are in the southeastern jet, and the extended SW structure represents the southwestern jet (Fig.~\ref{fig1}a-b). The brightest VLBI component W1 is possibly a jet knot in the proximity of the nucleus. This model naturally explains the long extended tongue, avoiding the conflict between the whole source age and the tongue age (as discussed above). However, the projected radio structure shows a large angle of about 90\degr{} between the two jet sides. 
Physical mechanisms responsible for such a sharp jet bend include jet--cloud interactions and pressure asymmetry in the NLR, but these interpretations require more data ({\it e.g.}, VLBI polarization data, and/or high-spatial-resolution spectroscopic data) to be confirmed or rejected.

{\em (3) A one-sided core-jet structure.}
A third distinct possibility is that the obscured core resides at the eastern end of the radio structure, i.e. in the component E. This makes the source a one-sided core--jet structure. One-sided jet structure in AGN is conventionally interpreted as a result of Doppler boosting. 
Therefore, this requires a highly relativistic jet and a small viewing angle to account for the absence of the counter jet. However, a small viewing angle seems inconsistent with the optical classification of Type 2. Moreover, this model also requires an explanation of the bend SW component. 

Model (1) seems the most plausible at this stage. New VLBI observations at another frequency, e.g. 1.6 or 2.3~GHz, are required to identify the location of a flat-spectrum core providing crucial discriminatory information. Optical spectroscopic observations or radio polarization measurements would provide diagnostic clues whether there is strong interaction with the ambient ISM at the large jet bending, as suggested in Model (2). Model (3) seems the least attractive, unless the radio jet axis is significantly misaligned with respect to the axis of symmetry of the host galaxy.

\begin{figure}
\includegraphics[scale=0.4]{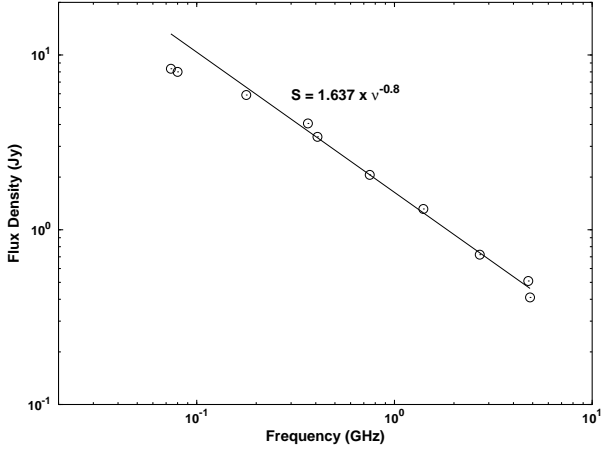}
\caption{Radio spectrum of 3C~316. The data points were obtained from the NED. The solid line represents the least-squares fit to the data at and above 160 MHz.}
\label{fig3}
\end{figure}

\subsection{Radio power characteristics} 

The total flux density measurements of 3C~316 at various radio frequencies were obtained from the NASA/IPAC Extragalactic Database\footnote{\url{http://ned.ipac.caltech.edu}} (NED), in order to investigate the integrated radio spectral index. The flux density data cover the frequency range 74~MHz--4.85~GHz at ten discrete frequencies (Fig.~\ref{fig3}). A least-square fit determines the power-law spectral index $\alpha$=0.80 ($S_\nu \propto \nu^{-\alpha}$) at and above 160~MHz, indicative of optically thin synchrotron emission. This is consistent with the radio imaging results where the jet and lobe dominate the total flux density. The flux densities at 74 and 80~MHz are lower than the fit by 58 and 55~per cent, respectively. This may indicate self-absorption in the source, and a low-frequency turnover between 80 and 160~MHz. 

The radio power of 3C\,316 may be calculated as:
\begin{equation}
P_{\rm 1.4} = 4\pi D^2_{\rm L} S_{\rm 1.4} (1+z)^{\alpha-1}, 
\end{equation}
where $S_{\rm1.4}$ is the flux density at 1.4~GHz, $D_{\rm L}$ the luminosity distance, $z$ the redshift, and $\alpha$ is the spectral index. This gives a radio power of $P_{\rm 1.4}$=1.7$\times$$10^{34}$~erg~s$^{-1}$Hz$^{-1}$, which places 3C~316 in the high-power MSO regime in the 'power--source size' diagram for extragalactic radio source evolution \citep{Kun10,AB12}.  Its source size ($\sim$3~kpc) suggests that 3C~316 is the beginning of the MSO phase, where the radio lobes just expanded out of the boundary between the ISM and the intergalactic medium (IGM). 
A radio source can evolve beyond this point if the radio lobes remain over-pressured with respect to the ambient medium and the jet remains powerful and supersonic, as manifested by well-defined, edge-brightened lobes \citep{KB07}.
However, the radio morphology of 3C~316 already shows signs of perturbations in the jet flow: it lacks well-confined edge-brightened lobes, it shows a series of embedded jet knots along an S-shaped trajectory, and it shows a long deflected outflow from a lobe (model 1), or large jet bend (model 2). 
This type of morphology is classified as the morphological Class-I source \citep{AB12}, which can maintain a long-term nuclear activity (i.e. providing continuous energy supply to the lobes over a time scale of $10^7$--$10^8$~yr) and may evolve into a large-scale FR I-like radio galaxy. However, if the nucleus ceases the supply of momentum flux, or the jet mechanical power is significantly reduced due to jet--ISM interactions, the radio source will not grow to Mpc scale. It would become a radio relic and the lobe will exist at low frequencies ($\lesssim$$10^2$~MHz) for a certain time.

\subsection{Black hole mass and accretion mode}

A tight correlation has been found between the central black hole mass ($M_{\rm BH}$) and the bulge velocity dispersion ($\sigma_*$) of the host galaxy \citep{Geb00,FM00}:
\begin{equation}
\log (M_{\rm BH}/M_\odot) = A + B \log (\sigma_*/\sigma_0),
\label{eq:M-sigma}
\end{equation}
where $\sigma_0$ is a reference value, usually taken as 200~km~s$^{-1}$.
\citet{Tre02} determined the parameters of the $M_{\rm BH}$--$\sigma_*$ relation as $A$=8.13$\pm$0.06 and $B$=4.02$\pm$0.32.
The difficulty of direct measurements of the bulge stellar velocity dispersion requires the use of a secondary estimate of the velocity dispersion of the ionized NLR gas. 
The [O\,{\sc III}] $\lambda$5007 is the strongest emission line in AGNs, and the velocity dispersion may be determined as $\sigma_*$=$\Delta V_{\rm [O\,III]}/2.35$.
Using this method, \citet{Nel00} found a strong correlation between $M_{\rm BH}$ and the [O\,{\sc III}] $\lambda$5007 line width, confirming the feasibility of using the gas velocity dispersion to represent the stellar velocity dispersion.

For 3C\,316 the use of the width of [O\,{\sc III}] as a measure of velocity dispersion is not straight-forward, given the complex structure of its line profile with two narrow peaks and a broad component, which is usually removed before estimating velocity dispersion. Therefore, in a very rough order-of-magnitude approach the width of the broad-wing component may be used as a {\em upper} limit on velocity dispersion and thus on the BH mass. 
Substituting this $\Delta V_{\rm [O\,III]}$=630.7~km~s$^{-1}$ (Table~\ref{tab2}) into the $M_{\rm BH}$--$\sigma_*$ relationship (Eq.~\ref{eq:M-sigma}),  a black hole mass for 3C\,316 is obtained in the range of 4.4$\times$$10^8$$M_\odot$.
When using $\Delta V_{\rm [O\,III]}$=99~km~s$^{-1}$ of the narrowest line component, an (implausible) lower limit of $2.6\times 10^5$ M$_\odot$ is obtained for the BH mass.

The Eddington luminosity of an accreting BH is defined as \citep{RL79}:
\begin{equation}
L_{\rm Edd} = \frac{4\pi G M_{\rm BH} m_{\rm p} c}{\sigma_{\rm T}} = 1.38 \times 10^{38} \left(\frac{M_{\rm BH}}{M_{\odot}}\right) {\rm erg}~{\rm s}^{-1} 
\end{equation}
where $\sigma_{\rm T}$ is the Thomson cross-section, $G$ the gravitational constant, $M_{\rm BH}$ the black hole mass, $c$ the light speed and $m_{\rm p}$ the proton mass.
For 3C\,316, $L_{\rm Edd}$ = 3.6$\times$$10^{43}$--6.1$\times$$10^{46}$~erg~s$^{-1}$.
In Seyfert 2 AGN, the bolometric luminosity based on a [O\,{\sc III}] line luminosity ($L_{\rm [O\,III]}$) requires an apriori-known dust extinction correction that is luminosity-dependent and is a factor of 454 for type 2 AGN with $L_{\rm [O\,III]}$=$10^{42}$--$10^{44}$ erg s$^{-1}$\citep{Lam09}.  This gives an extinction-corrected bolometric luminosity of $L_{\rm bol}$=1.4$\times$$10^{46}$ erg s$^{-1}$ for the 3C\,316 host galaxy. 

Using these estimates, the upper limit for BH mass suggests an Eddington ratio $\lambda_{\rm Edd}$=$L_{\rm bol}/L_{\rm Edd}$ = 0.23, which is consistent with the statistical mean value of $\lambda_{\rm Edd}$ = 0.28 for the compact steep-spectrum (CSS) sources with supposedly single AGN \citep{Wu09}. 
On the other hand, the lower limit for the BH mass gives a ratio of $\lambda_{\rm Edd}\sim$400, which would indicate that the BH is accreting close to its physical limit. Both values would imply high accretion activity during the early evolutionary phase of extragalactic radio sources and would exclude any low accretion rate models such as the advection-dominated accretion flow (ADAF) \citep{NY95}.   

\subsection{The NLR properties}

Because the evidence for binary black hole is not yet unambiguous,  the possibility of a kinematic explanation of the NLR properties needs to be investigated.  
The simplest NLR model is based on a single cloud of gas ionised as a central Str\"{o}mgren sphere, where the size of the NLR is correlated with the luminosity.
\citet{Sch03} found a relation between the NLR size and the [O\,{\sc III}] $\lambda$5007 luminosity: $R_{\rm maj}$$\propto$$L^{0.31}_{\rm [O\,III]}$ for a constant electron density. Using this relation and the [O\,{\sc III}] line luminosity of the broad-wing component, the NLR size of 3C\,316 is 2.2~kpc. However, NLR gas is always inhomogenuously distributed, and has a non-symmetric geometry. Long-slit optical spectroscopic observations found evidence of increasing gas density and ionization levels toward the nuclear source \citep{Ben02,Ben06}. In addition, the interactions between the radio jets and the NLR gas can generate shocks, which destroy the symmetry of the NLR gas and also enhance the local ionization. These factors together may result in a steeper slope of the $R$--$L_{\rm [O\,III]}$ relation and a larger NLR for 3C\,316. The estimated NLR size from the [O\,{\sc III}] luminosity is comparable to or slightly larger than the size of the radio structure, suggesting that the radio structure is still mostly confined within the NLR of the host galaxy or just expands out of the NLR. 

Given the electron density and temperature, the mass of the ionized gas in the NLR can be estimated from: 
\begin{equation}
M_{\rm H} = \frac{m_{\rm p} L_{\rm H\beta}}{\alpha_{\rm H\beta} n_{\rm e} h\nu_{\rm H\beta}},
\end{equation}
where $m_{\rm p}$ is the proton mass, $L_{\rm H{\beta}}$ the H$_{\rm \beta}$ luminosity, $\alpha_{\rm H\beta}$ the effective recombination coefficient for an H$_{\rm \beta}$ photon \citep[3$\times$$10^{-14}$\,cm$^3$s$^{-1}$;][]{Ost89}, and $h\nu_{\rm H\beta}$=4.09$\times$$10^{-12}$~erg the energy of an H${\rm \beta}$ photon. 
Assuming an electron density of $n_{\rm e}$=100~cm$^{-3}$ (e.g.), the mass of the NLR gas in 3C~316 is estimated to be $\sim$1.6$\times$$10^8$\,M$_\odot$.
This electron density is a typical value for the AGN NLR, and may disregard the more complex geometry and any clumpiness of the NLR, owing to the AGN-driven outflows, nuclear starbursts, winds driven by jet--cloud-induced bow shocks.  In young radio-loud AGN, the electron density $n_{\rm e}$ in the discrete clouds in the nuclear region can be even higher than $10^3$~cm$^{-3}$ \citep[$n_{\rm e}$$>$4$\times$$10^3$~cm$^{-3}$ in PKS~1345+12,][]{Holt03,Holt11}. However, the filling factors of such high-density clouds is small, and they may not contribute much to the total gas mass.
In addition, there is increasing evidence that the electron density in NLRs decreases with the radius \citep{Holt03,Holt06}. 
For instance, in NGC~1386 \citep{Ben06}, the electron density is 1000--1500~cm$^{-3}$ within 50~pc from the galactic nucleus, and drops  to 50~cm$^{-3}$ in the outskirts of the NLR (radius $>$3~kpc).
Due to the large uncertainty in the electron density ($n_{\rm e}$=$10^2$--$10^3$~cm$^{-3}$), the mass estimate of the NLR gas can vary by a factor of 10.

The H$\beta$ line may be used as an indicator of the rest-frame (systemic) velocity of the galaxy. The other weaker lines are shifted slightly relative to the H$\beta$ line and have slightly different velocity widths ($\Delta V_{\rm FWHM}$) (Table~\ref{tab2}). The broad-wing [O\,{\sc III}] component shows a blueshift relative to H$\beta$, suggesting an outflow of the emitting gas. The center velocities of the two narrow [O\,{\sc III}] components are also not symmetric relative to H$\beta$. 
Significant kinetic momentum is required to drive such outflow velocities. Assuming that the entire gas in the NLR is outflowing, adopting the  [O\,{\sc III}] velocity and the gas mass of 1.6$\times$$10^8$$M_\odot$, the kinetic momentum of the outflowing broad-wing gas is $M_{\rm H} v$=6.0$\times$$10^{48}$~g~cm~s$^{-1}$.
Given a typical AGN lifetime of $10^7$ yr, the time-averaged momentum injection rate is estimated to be 1.9$\times$$10^{34}$~g~cm~s$^{-2}$.
On the other hand, the radiation pressure of the AGN follows from the bolometric luminosity as $\tau (L_{\rm bol}/c)$=4.7$\times$$10^{35}$$\tau$, where $\tau$ is the effective optical depth.
Thus the injection rate of the kinetic momentum of the NLR gas is only $\sim$$(\frac{4.0}{\tau})$ per cent of the radiation pressure.  
As long as the opacity is not lower than 0.04, the radiation pressure of the AGN is indeed sufficient to drive the ionized outflow.

\subsection{The double-peaked narrow emission lines}

The two narrow [O\,{\sc III}] line components show line widths of 16 per cent and 27 per cent of that of the broad component.  They also show an asymmetric velocity structure relative to the galaxy rest frame with the blue narrow component having a higher relative velocity.
If these [O {\sc III}] line systems belong to a single AGN, the fact that both red- and blue-shifted components are detected may indicate a different origin from that of the broad-wing component. 
A straightforward first interpretation suggests that the narrow-peaked velocity components are associated with discrete gas clouds driven by bipolar outflows \citep{Hec84}. Alternatively, the dual peaks result from rotation in  a rotating disc/ring \citep{GH05} or a (partially self-absorbed) NLR sphere. 

The powerful radio jets in 3C\,316 naturally affect the NLR kinematics at large distances from the nucleus, and may provide mechanical feedback to the host galaxy in the form of high-velocity neutral and ionized outflows \citep[e.g.][]{Mor05,Holt03,Holt06}. Similarly, the high-velocity (491~km s$^{-1}$) [O\,{\sc III}] line component in the well-known quasar 3C~48 \citep{Sto07}, that is spatially associated with the radio jet \citep{Wil91,Wor04,Feng05,An10}, serves as evidence that NLR clouds can be affected by bow shocks of an expanding radio cocoon. 
The radio morphology of 3C\,316 could also show the signature of strong interactions with the NLR clouds, e.g., the knotty jet structure, the bright discrete jet knots as evidence of shock-brightening and jet-ISM interactions, and a curved jet trajectory signifying deflection. The shocks generated by these jet--cloud interactions provide secondary ionisation. 

Taking into account the limiting viewing angle of the jets and the observed velocity of the narrow [O\,{\sc III}] line components, the cocoon outflow velocity would be higher than 800~km~s$^{-1}$.   Besides the radiation-pressure-driven winds accounting for the broad-wing component, the radio jet may thus also provide additional acceleration of the clouds. 

A final question relates to the available accretion power that is being channeled into the radio jet.
The kinetic power of an FR I may be derived from a relation between the jet kinetic power $Q_{\rm jet}$ (in the form of X-ray cavity power) and the radiative power $P_{\rm jet}$ \citep{Bir08}, and the $Q_{\rm jet}$  of an FR II can be estimated directly from the jet terminal hot spot parameters \citep{GS13}.
Given a flux density of the 3C\,316 jet of $S_{\rm 1.4GHz} = 1.3$ Jy,  it gives a jet kinetic power of $Q_{\rm jet}$$\sim$$10^{45}$~erg~s$^{-1}$ using the $Q_{\rm jet}$--$P_{jet}$ relation given in Eq. (16) in \cite{Bir08}. 
A higher $Q_{\rm jet} \sim 2.5 \times 10^{45}$~erg~s$^{-1}$ is obtained by using Eq. (15) in \cite{Bir08} and $S_{\rm 327MHz} = 4.2$ Jy. 
This higher value of $Q_{\rm jet}$ is consistent with the estimate using the $Q_{\rm jet}$--$P_{\rm jet}$ relation for FR IIs using the Eq. (11) in \cite{GS13}, that gives $Q_{\rm jet} \sim 2.8 \times 10^{45}$ erg s$^{-1}$.
These $Q_{\rm jet}$s place 3C~316 in the mid-range of the CSS jet powers \citep{Wu09} as calculated from the low-frequency radio power \citep{Wil99}. 
The ratio of the jet power and the accretion power of 3C\,316 is then $Q_{\rm jet} / L_{\rm bol} = 0.07 - 0.20$, which means that about (7--20) per cent of the accretion power would be transferred into the radio jets. 
Quantitative modeling confirms that such a fraction (5--10 per cent) of accretion energy of the AGN can account for the mechanical feedback to the ISM of the host galaxy \citep{Tre02,DSH05}.  

Further improvement of these qualitative calculations to support scenario of jet--cloud interactions and to account for the double-peaked line profiles requires more knowledge of the size and mass distribution of the ionized clouds in the NLR. High-resolution optical images are necessary to identify the location of the red- and blue-shifted narrow velocity components. Polarimetric VLBI images are also required to verify the spatial correlation between the optical narrow-velocity components and the sites of jet-ISM interactions.

\section{Summary and conclusion}
\label{sec5}

\begin{enumerate}
\item High-resolution radio interferometric images of the prominent radio galaxy 3C\,316 have been presented for the first time. The images made from the EVN, e-MERLIN and VLA data (Fig.~\ref{fig1}) reveal the radio morphology on a variety of size scales. They consistently show a coherent radio structure made up by the Eastern, Central and Western components. A collimated diffuse structure starts from the Western component and extends south-westward to a distance of $\sim$2~kpc. The three major components are further resolved into a series of discrete compact jet components in the EVN image, resembling the knots in radio jets. The brightness temperatures of these knots range from 5.1$\times$$10^6$~K to 42.4$\times$$10^6$~K, but none of these VLBI components unambiguously identifies the AGN core. 

\item The observed radio structure tends to be consistent with the presence of a single radio-loud AGN in 3C~316, although the possibility of a (kpc-separation) radio quiet AGN cannot entirely be ruled out.  Indeed, the number of radio-confirmed binary SMBHs or dual AGN remains very small, which may suggest that most dual AGN are weak radio sources. The latest successful VLBI observations of a compact dual AGN system showed component flux densities below 1~mJy \citep[e.g. ][]{BP10,Frey12}. Alternatively, binary SMBHs only spend a relatively short time at kpc-scale separations during their evolution \citep{Burk11}.

\item The most prominent features in the SDSS spectrum of 3C~316 are the [O\,{\sc III}] emission lines with a double-peaked profile. Other narrow lines (e.g., H${\beta}$, [O\,{\sc II}]) do not show a double-peaked profile, because the weak fine structures are blended together. A broad-wing [O\,{\sc III}] line component shows a blueshift ($V_c = 188$ km s$^{-1}$ and $V_{\rm FWHM}$=600~km~s$^{-1}$), indicating large-scale outflowing motions.

\item The integrated flux densities of 3C\,316 shows a steep power-law spectrum ($\alpha$ = 0.8) from 160~MHz to 5~GHz. The projected linear size of the radio source is about 3~kpc. These characteristics classify 3C\,316 as an MSO or a CSS. 

\item The optical spectroscopic data of 3C\,316 imply an upper limit for the BH mass of 4.4$\times$$10^8$$M_\odot$ and a (very) lower limit of $2.6\times 10^5$$M_\odot$ adopting different values of the [O {\sc III}] line widths. The inferred Eddington ratio is higher than 0.23, which indicates that during the early evolutionary stages of the radio source the accretion activity is in a high state.

\item The broad-wing and narrow-velocity [O\,{\sc III}] components show differences in both velocity dispersion and velocity shift, which suggests that they may have different origins. The broad-wing component may be driven by the radiation pressure of the central ionizing source. The narrow-velocity components may be ionized locally and accelerated by the bow shocks of the radio jets.   
  
\end{enumerate}

In summary, the present observations of 3C\,316 did not provide any strong evidence for binary SMBHs. Follow-up radio observations are required to identify one (or more) core among the many radio knots. 
Furthermore, the interaction of the ISM with the radio jets is a plausible cause for the double-peaked [O\,{\sc III}] lines. This also makes this source of great interest for investigating the radio jet interactions with NLR gas, and the AGN feedback to the host galaxy. Further study is required in the following areas:
\begin{enumerate}
\item[(a)] Continuum VLBI imaging observations at e.g. 1.6~GHz would add the missing radio spectral index information, which discriminates between the AGN core and jet knot interpretation of the compact knots. 

\item[(b)] Polarimetric observations with the e-MERLIN at 5~GHz may provide direct evidence of jet--cloud interactions. Supplementary information of the neutral gas flows and the mechanical feedback may come from H\,{\sc I} absorption observations that can be done with the ASKAP or MeerKAT.  

\item[(c)] Further optical imaging with sub-arcsecond resolution is needed to determine the spatial correlation between the two optical components and the major radio components, to firmly confirm or rule out a dual AGN scenario.

\item[(d)] Optical long-slit spectroscopic observations are essential for determining the location and size of the [O\,{\sc III}] and H$\beta$ line components and for understanding the NLR kinematics including the physical properties of the line-emitting clouds and the outflow rates. Such quantitative measures strengthen our general understanding of the AGN feedback. 
\end{enumerate}

\section*{Acknowledgments}
\label{ack}
We thank the anonymous referee for his/her constructive comments which helped to improve the paper. 
This work was partly supported by the National Natural Science Foundation of Science and Technology of China (2009CB24900, 2013CB837900), the Strategic Priority Research Program on Space Science of the Chinese Academy of Sciences (CAS, No. XDA04060700), the Royal Dutch Academy of Sciences (KNAW) and the CAS exchange program (10CDP005), the China--Hungary Exchange Program of the CAS, the Hungarian Scientific Research Fund (OTKA K104539) and the NSFC (U1231115). 
T.~An has been supported by a visitor grant from the Netherlands Science Foundation (NWO No. 040.11.218).
T.~An thanks Cheng Li for contribution in initiating the EVN proposal, and thanks Luis C. Ho for helpful comments on the manuscript.
Y.-H.~Xu thanks for the hospitality of JIVE. 
We thank the e-MERLIN operations team for data taken during commissioning time.
The EVN is a joint facility of European, Chinese, South African and other radio astronomy institutes funded by their national research councils. 
The e-MERLIN is a National Facility operated by the University of Manchester at Jodrell Bank Observatory on behalf of the UK Science and Technology Facilities Council (STFC). 
This work has benefited from research funding from the European Community's sixth Framework Programme under RadioNet R113CT 2003 5058187.
The National Radio Astronomy Observatory is a facility of the National Science Foundation operated under cooperative agreement by Associated Universities, Inc. 
This research has made use of the NASA/IPAC Extragalactic Database (NED) which is operated by the Jet Propulsion Laboratory, California Institute of Technology, under contract with the National Aeronautics and Space Administration.

\label{lastpage}
\end{document}